\documentclass[reqno,12pt,a4paper]{amsart}

\usepackage{mathrsfs}
\usepackage{amssymb}
\usepackage{amsfonts}
\usepackage{latexsym}
\usepackage{amsthm}

\usepackage{graphicx}
\def\lb{\label}

\newcommand{\er}[1]{\textrm{(\ref{#1})}}

\begin{document}


\renewcommand{\theequation}{\arabic{section}.\arabic{equation}}
\theoremstyle{plain}
\newtheorem{theorem}{\bf Theorem}[section]
\newtheorem{lemma}[theorem]{\bf Lemma}
\newtheorem{corollary}[theorem]{\bf Corollary}
\newtheorem{proposition}[theorem]{\bf Proposition}
\newtheorem{definition}[theorem]{\bf Definition}
\newtheorem{remark}[theorem]{\it Remark}

\def\a{\alpha}  \def\cA{{\mathcal A}}     \def\bA{{\bf A}}  \def\mA{{\mathscr A}}
\def\b{\beta}   \def\cB{{\mathcal B}}     \def\bB{{\bf B}}  \def\mB{{\mathscr B}}
\def\g{\gamma}  \def\cC{{\mathcal C}}     \def\bC{{\bf C}}  \def\mC{{\mathscr C}}
\def\G{\Gamma}  \def\cD{{\mathcal D}}     \def\bD{{\bf D}}  \def\mD{{\mathscr D}}
\def\d{\delta}  \def\cE{{\mathcal E}}     \def\bE{{\bf E}}  \def\mE{{\mathscr E}}
\def\D{\Delta}  \def\cF{{\mathcal F}}     \def\bF{{\bf F}}  \def\mF{{\mathscr F}}
\def\c{\chi}    \def\cG{{\mathcal G}}     \def\bG{{\bf G}}  \def\mG{{\mathscr G}}
\def\z{\zeta}   \def\cH{{\mathcal H}}     \def\bH{{\bf H}}  \def\mH{{\mathscr H}}
\def\e{\eta}    \def\cI{{\mathcal I}}     \def\bI{{\bf I}}  \def\mI{{\mathscr I}}
\def\p{\psi}    \def\cJ{{\mathcal J}}     \def\bJ{{\bf J}}  \def\mJ{{\mathscr J}}
\def\vT{\Theta} \def\cK{{\mathcal K}}     \def\bK{{\bf K}}  \def\mK{{\mathscr K}}
\def\k{\kappa}  \def\cL{{\mathcal L}}     \def\bL{{\bf L}}  \def\mL{{\mathscr L}}
\def\l{\lambda} \def\cM{{\mathcal M}}     \def\bM{{\bf M}}  \def\mM{{\mathscr M}}
\def\L{\Lambda} \def\cN{{\mathcal N}}     \def\bN{{\bf N}}  \def\mN{{\mathscr N}}
\def\m{\mu}     \def\cO{{\mathcal O}}     \def\bO{{\bf O}}  \def\mO{{\mathscr O}}
\def\n{\nu}     \def\cP{{\mathcal P}}     \def\bP{{\bf P}}  \def\mP{{\mathscr P}}
\def\r{\rho}    \def\cQ{{\mathcal Q}}     \def\bQ{{\bf Q}}  \def\mQ{{\mathscr Q}}
\def\s{\sigma}  \def\cR{{\mathcal R}}     \def\bR{{\bf R}}  \def\mR{{\mathscr R}}
\def\S{\Sigma}  \def\cS{{\mathcal S}}     \def\bS{{\bf S}}  \def\mS{{\mathscr S}}
\def\t{\tau}    \def\cT{{\mathcal T}}     \def\bT{{\bf T}}  \def\mT{{\mathscr T}}
\def\f{\phi}    \def\cU{{\mathcal U}}     \def\bU{{\bf U}}  \def\mU{{\mathscr U}}
\def\F{\Phi}    \def\cV{{\mathcal V}}     \def\bV{{\bf V}}  \def\mV{{\mathscr V}}
\def\P{\Psi}    \def\cW{{\mathcal W}}     \def\bW{{\bf W}}  \def\mW{{\mathscr W}}
\def\o{\omega}  \def\cX{{\mathcal X}}     \def\bX{{\bf X}}  \def\mX{{\mathscr X}}
\def\x{\xi}     \def\cY{{\mathcal Y}}     \def\bY{{\bf Y}}  \def\mY{{\mathscr Y}}
\def\X{\Xi}     \def\cZ{{\mathcal Z}}     \def\bZ{{\bf Z}}  \def\mZ{{\mathscr Z}}
\def\O{\Omega}
\def\vr{\varrho}
\def\vs{\varsigma}

\newcommand{\gA}{\mathfrak{A}}          \newcommand{\ga}{\mathfrak{a}}
\newcommand{\gB}{\mathfrak{B}}          \newcommand{\gb}{\mathfrak{b}}
\newcommand{\gC}{\mathfrak{C}}          \newcommand{\gc}{\mathfrak{c}}
\newcommand{\gD}{\mathfrak{D}}          \newcommand{\gd}{\mathfrak{d}}
\newcommand{\gE}{\mathfrak{E}}
\newcommand{\gF}{\mathfrak{F}}           \newcommand{\gf}{\mathfrak{f}}
\newcommand{\gG}{\mathfrak{G}}           
\newcommand{\gH}{\mathfrak{H}}           \newcommand{\gh}{\mathfrak{h}}
\newcommand{\gI}{\mathfrak{I}}           \newcommand{\gi}{\mathfrak{i}}
\newcommand{\gJ}{\mathfrak{J}}           \newcommand{\gj}{\mathfrak{j}}
\newcommand{\gK}{\mathfrak{K}}            \newcommand{\gk}{\mathfrak{k}}
\newcommand{\gL}{\mathfrak{L}}            \newcommand{\gl}{\mathfrak{l}}
\newcommand{\gM}{\mathfrak{M}}            \newcommand{\gm}{\mathfrak{m}}
\newcommand{\gN}{\mathfrak{N}}            \newcommand{\gn}{\mathfrak{n}}
\newcommand{\gO}{\mathfrak{O}}
\newcommand{\gP}{\mathfrak{P}}             \newcommand{\gp}{\mathfrak{p}}
\newcommand{\gQ}{\mathfrak{Q}}             \newcommand{\gq}{\mathfrak{q}}
\newcommand{\gR}{\mathfrak{R}}             \newcommand{\gr}{\mathfrak{r}}
\newcommand{\gS}{\mathfrak{S}}              \newcommand{\gs}{\mathfrak{s}}
\newcommand{\gT}{\mathfrak{T}}             \newcommand{\gt}{\mathfrak{t}}
\newcommand{\gU}{\mathfrak{U}}             \newcommand{\gu}{\mathfrak{u}}
\newcommand{\gV}{\mathfrak{V}}             \newcommand{\gv}{\mathfrak{v}}
\newcommand{\gW}{\mathfrak{W}}             \newcommand{\gw}{\mathfrak{w}}
\newcommand{\gX}{\mathfrak{X}}               \newcommand{\gx}{\mathfrak{x}}
\newcommand{\gY}{\mathfrak{Y}}              \newcommand{\gy}{\mathfrak{y}}
\newcommand{\gZ}{\mathfrak{Z}}             \newcommand{\gz}{\mathfrak{z}}

\def\be{{\bf e}} \def\bc{{\bf c}}
\def\bv{{\bf v}} \def\bu{{\bf u}}
\def\mm{\mathrm m}
\def\mn{\mathrm n}

\def\ve{\varepsilon}   \def\vt{\vartheta}    \def\vp{\varphi}    \def\vk{\varkappa}

\def\Z{{\mathbb Z}}    \def\R{{\mathbb R}}   \def\C{{\mathbb C}}\def\K{{\mathbb K}}
\def\T{{\mathbb T}}    \def\N{{\mathbb N}}   \def\dD{{\mathbb D}} \def\dP{{\mathbb P}}
\def\B{{\mathbb B}}


\def\la{\leftarrow}              \def\ra{\rightarrow}      \def\Ra{\Rightarrow}
\def\ua{\uparrow}                \def\da{\downarrow}
\def\lra{\leftrightarrow}        \def\Lra{\Leftrightarrow}


\def\lt{\biggl}                  \def\rt{\biggr}
\def\ol{\overline}               \def\wt{\widetilde}
\def\no{\noindent}               \def\ti{\tilde}
\def\ul{\underline}


\let\ge\geqslant                 \let\le\leqslant
\def\lan{\langle}                \def\ran{\rangle}
\def\/{\over}                    \def\iy{\infty}
\def\sm{\setminus}               \def\es{\emptyset}
\def\ss{\subset}                 \def\ts{\times}
\def\pa{\partial}                \def\os{\oplus}
\def\om{\ominus}                 \def\ev{\equiv}
\def\iint{\int\!\!\!\int}        \def\iintt{\mathop{\int\!\!\int\!\!\dots\!\!\int}\limits}
\def\el2{\ell^{\,2}}             \def\1{1\!\!1}
\def\wh{\widehat}

\def\sh{\mathop{\mathrm{sh}}\nolimits}
\def\ch{\mathop{\mathrm{ch}}\nolimits}

\def\where{\mathop{\mathrm{where}}\nolimits}
\def\as{\mathop{\mathrm{as}}\nolimits}
\def\Area{\mathop{\mathrm{Area}}\nolimits}
\def\arg{\mathop{\mathrm{arg}}\nolimits}
\def\const{\mathop{\mathrm{const}}\nolimits}
\def\det{\mathop{\mathrm{det}}\nolimits}
\def\diag{\mathop{\mathrm{diag}}\nolimits}
\def\diam{\mathop{\mathrm{diam}}\nolimits}
\def\dim{\mathop{\mathrm{dim}}\nolimits}
\def\dist{\mathop{\mathrm{dist}}\nolimits}
\def\Im{\mathop{\mathrm{Im}}\nolimits}
\def\Iso{\mathop{\mathrm{Iso}}\nolimits}
\def\Ker{\mathop{\mathrm{Ker}}\nolimits}
\def\Lip{\mathop{\mathrm{Lip}}\nolimits}
\def\rank{\mathop{\mathrm{rank}}\limits}
\def\Ran{\mathop{\mathrm{Ran}}\nolimits}
\def\Re{\mathop{\mathrm{Re}}\nolimits}
\def\Res{\mathop{\mathrm{Res}}\nolimits}
\def\res{\mathop{\mathrm{res}}\limits}
\def\sign{\mathop{\mathrm{sign}}\nolimits}
\def\span{\mathop{\mathrm{span}}\nolimits}
\def\supp{\mathop{\mathrm{supp}}\nolimits}
\def\Tr{\mathop{\mathrm{Tr}}\nolimits}
\def\BBox{\hspace{1mm}\vrule height6pt width5.5pt depth0pt \hspace{6pt}}


\newcommand\nh[2]{\widehat{#1}\vphantom{#1}^{(#2)}}
\def\dia{\diamond}

\def\Oplus{\bigoplus\nolimits}



\def\qqq{\qquad}
\def\qq{\quad}
\let\ge\geqslant
\let\le\leqslant
\let\geq\geqslant
\let\leq\leqslant
\newcommand{\ca}{\begin{cases}}
\newcommand{\ac}{\end{cases}}
\newcommand{\ma}{\begin{pmatrix}}
\newcommand{\am}{\end{pmatrix}}
\renewcommand{\[}{\begin{equation}}
\renewcommand{\]}{\end{equation}}
\def\bu{\bullet}


\title[{Trace formulas for time periodic complex Hamiltonians  }]
 {Trace formulas for time periodic complex Hamiltonians on lattice}

\author[Evgeny, L. Korotyaev]{Evgeny, L. Korotyaev}
\address{E. Korotyaev, Depart. of Math. Analysis, Saint-Petersburg State University,
Universitetskaya nab. 7/9, St. Petersburg, 199034, Russia, \
korotyaev@gmail.com, \ e.korotyaev@spbu.ru}

\date{\today}

\begin{abstract}
\no  We consider time periodic Hamiltonians with complex potentials
on the lattice and determine trace formulas. As a corollary we
estimate  eigenvalues of the quasienergy operator in terms  of the
norm of potentials.

\end{abstract}


\maketitle

\section {Introduction and main results}\label{sec:1}
\setcounter{equation}{0}

\subsection{Introduction}

We discuss scattering and trace formulas for the Schr\"odinger
equation on the lattice $\Z^d$:
\[
\lb{01}
\begin{aligned}
{\tfrac{ d }{d t}}u(t)= -i h(t) u(t),\qq h(t)=\D+ V(t),
\end{aligned}
\]
where $h(t) $ is the Hamiltonian,  $\t$-periodic in time $t$ and
$\D$ is the discrete Laplacian given by
\[
\big(\D f\big)_x=\frac{1}{2}\sum_{|x-y|=1}(f_x- f_y), \qqq
f=(f_x)_{x\in{\Z}^d} \in \ell^{2}({\Z}^d),\qq  x=(x_j)_1^d\in\Z^d.
\]
It is known that the spectrum of the Laplacian $\D$ is absolutely
continuous and satisfies
$$
\s(\D)=\s_{\textup{ac}}(\D)=[0,2d].
$$
Here $V(t)$ is $\t$-periodic in time potential: $(V(t)
f)_x=V_x(t)f_x,$ for all $(t,x)\in \R\ts \Z^d$.
 Introduce the space $\ell^p(\Z^{d}), p\ge 1$  of
sequences $f=(f_x)_{x\in \Z^d}$ equipped with the norm
 given by
$$
\begin{aligned}
\|f\|_{p}=\|f\|_{\ell^p(\Z^{d})} =\big(\sum_{x\in
\Z^d}|f_x|^p\big)^{1\/p},\qq  \ p\in [1,\iy),
\end{aligned}
$$
and let $\|f\|_{\iy}=\|f\|_{\ell^\iy(\Z^{d})} =\sup_{x\in
\Z^d}|f_x|$. For a Banach space $\gB$ we write $\ell^r(\gB), r\ge 1$
for the space of $\gB$-valued sequences with $p^{th}$ power summable
norms, and $L^r(\T_\t,\gB)$ for the $\gB$-valued $L^r$-space. In the
case $f(\cdot)\in L^r(\T_\t,\ell^p(\Z^{d})$ we define  the norm
$\|f\|_{r,p}$ by
\[
\|f\|_{p,r}^r=\int_{\T_\t}\|f(t)\|_{\ell^p(\Z^{d})}^rdt, \qq p,r \ge
1.
\]
Note that $\|f\|_{p,r}\le \|f\|_{q,r}$ for all $p\ge q\ge 1$. We
assume that potentials can be complex-valued and  satisfy

{\bf Condition V.} {\it Let  $d\ge 3$.
  The function $V(t)$  is $\t$ -periodic, and satisfies }
\begin{equation}
\label{V}
\begin{aligned}
\|V\|_{p,2}<\iy,
\qqq\ca 1\le p<{6\/5} & if \ \ d=3\\
1\le p<{4\/3} & if  \  \ d\ge 4 \ac.
\end{aligned}
\end{equation}

We discuss trace formulas for operators with complex potentials.
Recall that, in general, a trace formula is an identity connecting
the integral of the potential and various sums of eigenvalues and
integrals of coefficients of S-matrix of the Schr\"odinger operator
(or other spectral characteristics). We shortly describe results
about multidimensional trace formulas:


$\bu $ Real potentials. The first result was obtained by Buslaev
\cite{B66}, see also \cite{G85}, \cite{P82} and references therein.
Trace formulas for Stark operators and magnetic Schr\"odinger
operators were discussed in \cite{KP03}, \cite{KP04}.
 The trace formulas for Schr\"odinger operators on the lattice
$\Z^d$ with real decaying potentials  were determined by
Isozaki--Korotyaev \cite{IK12}.

$\bu $  Complex potentials. Unfortunately, we know only few papers
about the trace formulas for Schr\"odinger operators with
complex-valued potentials decaying at infinity. Trace formulas for
Schr\"odinger operators with    complex decaying potentials were
determined by Korotyaev \cite{K20R} in the continuous case and in
the discrete case by Korotyaev and Laptev \cite{KL18}, Korotyaev
\cite{K17} and for the specific case $\Im V\le 0$ by Malamud and
Neidhardt \cite{MN15}.

 Our main goal is to determined trace formulas for time periodic
Hamiltonian with complex (and for real) potentials on the lattice.
The case of $\R^d$ is more complicated \cite{K21}. We do not know
any results about it.

\medskip

 For Hilbert space $\cH$ and $\T_\t=\R/(\t \Z)$ we introduce the space $\ti
\cH=L^2(\T_\t,\cH)$ of functions $f\to f(t)$ that are $\t$--periodic
in time with values in $\cH$ equipped with the norm
$$ \|f\|^2_{\ti
\cH}={1\/\t}\int_0^\t \|f(t)\|^2_{\cH}dt.
$$
The space $\ti \cH$ can be realized as $\ell^2(\cH)$ via the Fourier
transform $\F: \ti \cH\to \ell^2(\cH)$ defined by
$$
\begin{aligned}
f\to \F f=(f_n)_{n\in \Z}, \qq  f_n=(\F f)_n={1\/\sqrt \t}\int_0^\t
e^{-int\o}f(t)dt,\qq \o={2\pi\/\t},\qq  f\in \ti \cH.
\end{aligned}
$$
Let $\pa=-i\frac{\pa}{\pa t}$ be the  self-adjoint operator in
$L^2(\T_\t)$. We also denote $\pa=-i\frac{\pa}{\pa t}$ the
corresponding operator in $\wt \cH$ with the natural domain
$\mD=\mD(\pa)$.  We use the notation  $\lan A(t)\ran$ to indicate
multiplication  by $A(t)$ on the space $\wt\cH$. Introduce the
operators $\wt h_o$ and $\wt h$ on $\wt\cH=L^2(\T_\t, \ell^2(\Z^d))$
by
$$
\wt h_o=\pa+\D,\qqq  \wt h=\wt h_o+\lan V(t) \ran.
$$
It is known that the spectrum $\s(\D)=[0,2d]$. Then the spectrum of
$\wt h_o$ has the form
\[
\s(\wt h_o)=\s_{ac}(\wt h_o)=\bigcup_{n\in \Z} \s(\D+\o
n)=\bigcup_{n\in \Z} [\o n,\o n+2d].
\]
Note that if $\o >2d$, then the spectrum of $\wt h_o$ has the band
structure with the bands $\s(\D+\o n)=[\o n,\o n+2d]$ separated by
gaps.  Let $\cB_1$ and $\cB_2$ be the trace and the Hilbert-Schmidt
class equipped with the norm $\|\cdot \|_{\cB_1}$ and $ \|\cdot
\|_{\cB_2}$, respectively. Introduce the free resolvent $
R_o(\l)=(\wt h_o-\l)^{-1}, \l\in \C_\pm$. Below we show that if $V$
satisfies Condition V, then
\[
\lb{1.c} VR_o(\l)\in \cB_2, \qq \forall \ \l\in \C_\pm.
\]
This yields $\mD(\wt h)=\mD(\wt h_o)$ and
\[
\lb{1.c1}  \s_{ess}(\wt h)=  \s_{ess}(\wt h_o).
\]
Thus  the operator $\wt h$ has only discrete spectrum in $\C_\pm$.
Define the perturbed resolvent $R(\l)=(\wt h-\l)^{-1} $ for all
$\l\in \C_\pm\sm \s_{disc}(\wt h)$. We have the very useful identity
\[
\lb{i1} \lan e^{it\o}\ran R(\l)\lan e^{-it\o}\ran  =R(\l+\o),\qq
\forall \l\in \C_\pm.
\]
It means that the spectrum of $\wt h$ (and $\wt h_o$)  is
$\o$--periodic. Thus it is sufficient to study eigenvalues of $\wt
h$ in a  strip $\Re \l\in [0,\o)$. We consider the case of the half
strip $\L\ss \C_+$ defined by
$$
\textstyle \L=[0,\o)\ts i\R_+\ss \C_+, \qqq \o={2\pi\/\t}.
$$
The proof for the lower half strip  $\ol \L$ is similar. The
operator $\wt h$ has $N\le \iy$ eigenvalues $\{\l_j, j=1,....,N\}$
in the strip $\L$. Here and below each eigenvalue is counted
according to its algebraic multiplicity. We have similar
consideration for the case $\C_-$.

\subsection {Main results}

We assume that a potential $V$ satisfy Condition V. We define a
operator-valued function $\gF$ (below we show that $\gF(\l)\in
\cB_2$) and the regularized determinant $\cD$  by
\[
\lb{dgF} \gF(\l)=|V|^{1\/2}R_0(\l) |V|^{1\/2}e^{i\arg V},\qqq \qqq
\l \in \C_\pm,
\]
and
\[
\lb{De2}
\begin{aligned}
& \cD(\l)=\det \Big[(I+\gF)e^{-\gF}\Big](\l),\qqq \l\in\C_\pm.
\end{aligned}
\]
$\cD$ is the basic function to study trace formulas.  We describe
main properties of  $\cD$.

\begin{theorem}
\lb{T1}  Let $V$ satisfy Condition V and a constant $C_*$ be defined
by \er{Cpd}.

\no i)  Then the operator-valued function $\gF\colon \C_\pm \to
\cB_2(L^2(\T_\t,\ell^2(\Z^d)))$, defined by \er{dgF}  is analytic
and H\"older continuous up to the boundary. Moreover, it satisfies:
\[
\label{D1}
\begin{aligned}
 \|\gF(\l)\|_{\cB_2}\le \cC_\bu \|V\|_{p,2}\qqq \forall \ \l\in \C_\pm,
 \\
\cC_\bu=1+ (1+{\t d\/\pi}) \big(C_g+\t^{-{1\/2}} \big)+{C_*\/\t},
 \end{aligned}
\]
where $C_g={4\/3\pi}+{5+3\pi\/4}\sqrt2$ and the constant $ C_*$ is
defined by \er{Cpd}.

\no ii) The modified determinant $\cD$ is analytic in $\C_\pm$,
H\"older up to the boundary and satisfies
\[
\label{D2} \cD(\l+\o)=\cD(\l)\qqq \forall \ \l\in \C_\pm,
\]
\[
\label{D3} \cD(\l)=1+{O(1)\/\n} \qq \as \qq \n:=|\Im \l|\to \iy,
\]
\begin{equation}
\label{D4}
\begin{aligned}
\sup_{\l\in \C_\pm} |\cD(\l)|\le e^{{\cC_\bu^2\/2}\|V\|_{p,2}^2}.
\end{aligned}
\end{equation}
Moreover, if $\l\in \C\sm \R$ is an eigenvalue of  the operator $\wt
h$, then
\[
\lb{ee}  \n(1-e^{- \t\n})\le 2\|V\|_{2,2}^2.
\]

\end{theorem}

\no {\bf Remark.}  1) For complex potentials we discuss eigenvalues
of $\wt h$ only in the domain $[0,\o]\ts (0, i\n_o],
\n_o=2\|V\|_{2,2}^2$, since the operator $\wt h$ does not have zeros
in the domain $\{|\Im \l|>\n_o\}$.

2) If the operator $V$ is bounded, then eigenvalues of $\wt h$
belong to the strip $\{|\Im \l|<\|\Im V\| \}$.

\medskip

We define the disc $\dD_r\subset\C$ with the radius $r>0$   by $
\dD_r=\{z\in \C:|z|<r\}, $ and abbreviate  $\dD=\dD_1$. We define
the Hardy space in a disk $\dD$.  We say a function $F$ belongs the
Hardy space $ \mH=\mH_\iy(\dD)$ if $F$ is analytic in $ \dD$ and
satisfies
$$
\|F\|_{\mH}:=\sup_{\l\in \dD}|F(z)|<\iy.
$$
For $\l\in \L$  we define the new spectral variable $z\in \dD$ by
$$
\begin{aligned}
\textstyle
 z=e^{i \t\l}\in \dD,\qqq  \l(z)=-{i\/\t}\ln z\in
\L=[0,\o]\ts \R_+\ss \C_+.
\end{aligned}
$$
The function $z=e^{i \t\l}, \l\in\L$ is a conformal mapping from the
strip $\L$ onto the unit disk $\dD$. Theorem \ref{T1} shows that a
function
$$
\p(z):=\cD(\l(z)),\qqq  z\in \dD,
$$
 belongs to the Hardy space $\mH_\iy(\dD)$.

 Define  operators $J_1, J_2$ on $L^2(\T_\t)$ and operators $F_1, F_2$ on $\wt\cH$ by
\[
\begin{aligned}
\lb{op1i}
 (J_1f)(t)=i\int_0^t f(s)ds,\qqq (J_2 f)(t)= i\int_0^\t f(s)ds,
\end{aligned}
\]
and
\[
\begin{aligned}
\lb{dd2xx}
 F_1=J_1 \wt V,\qq  F_2=J_2\wt V, \qqq \wt
V(t)=e^{it\D}V(t)e^{-it\D},\qqq \cR_1 =(I+F_1)^{-1}.
\end{aligned}
\]
Note that  the operator $I+F_1$ is  invertible.

\begin{theorem}
\lb{T2} Let a potential $V$ satisfy Condition V and the constant
$C_\bu$ be defined by \er{D1}. Then the function $\p(z):=\cD(\l(z))$
belongs to $\mH_\iy(\dD)$ and is H\"older up to the boundary and
satisfies
\begin{equation}
\label{D12}
\begin{aligned}
\|\p\|_{\mH_\iy(\dD)}\le e^{{C_\bu^2\/2}\|V\|_{p,2}^2}.
\end{aligned}
\end{equation}
The zeros $\{z_j\}_{j=1}^N$ of  $\p$ in $\dD$ satisfy $
 \sum _{j=1}^N (1-|z_j|)<\iy$.  Moreover, the function $\log \p(z)$
is analytic in $\dD_{r_o}$ for some $r_o>0$ and has the Taylor
series :
\[
\label{D32}
\begin{aligned}
 \log \p(z)=\p_1z+\p_2z^2+\p_3z^3 +......, \qqq \as \qq |z|<r_o,
\end{aligned}
\]
where the coefficients $\p_n$ are given by
\[
\label{D42}
\begin{aligned}
 \p_1=-\Tr F_1 \cR_1 e_\t F_2 , \qqq \p_2= -\Tr F_1 \cR_1
e_\t^2 F_2+{1\/2}\Tr F_1 (\cR_1 e_\t F_2 )^2,....
\end{aligned}
\]
and $e_\t=e^{-i\t \D}$  and  $F_1, F_2$ are defined by \er{dd2xx}
and $\cR_1=(1+F_1)^{-1} $.

\end{theorem}

{\bf Remark.} 1) We transform the analytic problem  from the domain
$\C_+$ to the disk $\dD$. The energy periodic property \er{D2} of
the determinant $\cD(\l+\o)=\cD(\l)$ and asymptotics \er{D3} are
crucial here.

2) It is unusual that due to \er{D32} the determinant $\log
\cD(\l)=e^{-\t \Im \l}O(1)$ as $\Im \l\to \iy$.

\medskip

Recall that the operator $\wt h$ has $N\le \iy$ eigenvalues $\{\l_j,
j=1,....,N\}$ in the domain $\L$. Each point $z_j=z(\l_j)\in \dD$ is
a zero of $\p(z)$. For the function $\p$ we define the Blaschke
product $B(z), z\in \dD$ by: $B=1$ if $N=0$ and
\[
\lb{B2}
\begin{aligned}
 B(z)=\prod_{j=1}^N {|z_j|\/z_j}{(z_j-z)\/(1-\ol z_j z)},\qqq
 z_j=e^{i \t\l_j}\qqq
  if \qqq N\ge 1.
\end{aligned}
\]
It is well known that the Blaschke product $B(z), z\in \dD$ given by
\er{B2} converges absolutely for $\{|z|<1\}$ and  satisfies $B\in
\mH_\iy(\dD)$ with $\|B\|_{\mH_\iy}\le 1$, since $\p\in \mH_\iy$
\cite{G81}. The Blaschke product $B$ has the standard Taylor series
at $z=0$:
\[
\begin{aligned}
\lb{B6d} & \log  B(z)=B_0-B_1z-B_2z^2-... \qqq as \qqq z\to 0,
\end{aligned}
\]
where $B_0=\log  B(0)<0$ and
$B_n={1\/n}\sum_{j=1}^N\Big({1\/z_j^n}-\ol z_j^n \Big), n\ge 1$. In
particular we have
\[
\begin{aligned}
\lb{BL}
 B_0=\log  B(0)=-\t\sum\Im \l_j<0.
\end{aligned}
\]

\noindent We describe the canonical representation of the
determinant $\p(z), z\in\dD$.

\begin{corollary}
\lb{T3} Let a potential $V$ satisfy Condition V. Then the
determinant $\p$ has a canonical factorization for all $|z|<1$ given
by
\[
\lb{cfD}
\begin{aligned}
& \p(z)=B(z) e^{\P(z)},\qq
 \P(z)={1\/2\pi}\int_{0}^{2\pi}{e^{it}+z\/e^{it}-z}d\m(t),
 \\
 & d\m(t)=\ln |\p(e^{it})|dt-d\gm(t),
 \end{aligned}
\]
 where $\ln |\p(e^{it}) |\in L^1(\T)$ and $\gm\ge 0$ is some singular measure  on $[0,2\pi]$,
such that $ \supp \gm\ss \{t\in [0,2\pi]: \p(e^{it})=0\}$. Moreover,
$\P$ has the Taylor series at $z=0$ in some disk $\{|z|<r\}$:
\[
\lb{asmm} \P(z)={\m(\T)\/2\pi}+\m_1z+ \m_2z^2+\m_3z^3+\m_4z^4+...,
\]
where
$$
\m(\T)=\int_0^{2\pi}d\m(t)=\int_0^{2\pi}\log |\p(e^{it})|dt-\gm(\T),
\qqq \m_n={1\/\pi}\int_0^{2\pi}e^{-int} d\m(t),\qqq n\in \N.
$$

\end{corollary}

We present our main result about trace formulas.

\smallskip
\begin{theorem}
\lb{T4}   Let $V$ satisfy Condition V. Then
 the following trace formula holds true:
\[
\lb{trR} -{i\/\t z}\Tr \rt(R(\l)-R_o(\l) +R_o(\l)VR_o(\l) \rt)= \sum
{(1-|z_j|^2)\/(z-z_j)(1-\ol
z_jz)}+{1\/\pi}\int_{0}^{2\pi}{e^{it}d\m(t)\/(e^{it}-z)^2} ,
\]
\[
\lb{t1} {\gm(\T)\/2\pi}+\t\sum_{\l\in \L}\Im
\l_j={1\/2\pi}\int_{0}^{2\pi}\log |\p(e^{it})|dt\ge 0,
\]
\[
\lb{t2} B_n=\p_n+\m_n,\qqq n=1,2,3,....
\]
where $\l(z)={i\/\t}\ln z,\ z\in \dD$ and the measure $d\m(t)=\log
|\p(e^{it})|dt-d\gm(t)$, and $B_n$ are given by \er{B6d},  and in
particular,
\[
\lb{t3} B_1=\sum_{j=1}^N\rt({1\/z_j}-\ol z_j
\rt)=\p_1+{1\/\pi}\int_\T e^{-it}d\m(t),
\]

\end{theorem}

{\bf Remark.} 1) The measure $d\m(t)$ in \er{trR} is some analog of
the spectral shift function for complex potentials.

2) The trace formula  \er{t1} has the term $\gm(\R)$ which is absent
for real potentials. There is an open problem: when this term is
absent (or there exists)  for specific complex potentials.

3) Consider the $ \p(z)=\cD(\l(z)),\  z\in \dD$. If $z=e^{it}$, then
$\l={\ln z\/i\t}={t\/\t} \in [0,\o]$. Then we obtain
$$
\int_{0}^{2\pi}\log |\p(e^{it})|dt=\t \int_{0}^{\o}\log
|\cD(\l+i0)|d\l.
$$
Moreover, we can do the same with all integrals in Theorem \ref{T4}.

\medskip

\begin{corollary}
\lb{T5} Let a potential $V$ satisfy Condition V. Then the following
estimate hold true:
\[
\lb{e1} {\gm(\T)\/2\pi}+\t\sum_{\l\in \L}\Im \l_j\le
{C_\bu^2\/2}\|V\|_{p,2}^2,
\]
where the constant $C_\bu$ is defined by \er{D1}.

\end{corollary}

{\bf Remark.} 1) The measure $d\m(t)$ in \er{trR} is some analog of
the spectral shift function for complex potentials.

2) If a potential $V$ does not depend on time then there are
estimates of complex eigenvalues in terms of potentials, see
\cite{KL18}, \cite{K17}. In the continuous case there are a lot of
results about it, see, e.g., \cite{FLS16}, \cite{F18} and references
therein.

For time-periodic Hamiltonians many papers have been devoted to
scattering  mainly for operators $h(t)=-\D + V(t,x)$ on $\R^d, d\ge
1$, and to the spectral analysis of the corresponding monodromy
operator. Zel'dovich \cite{Z73} and Howland \cite{H79} reduced
 the problem with a time-dependent Hamiltonian  to a problem with a
time-independent Hamiltonian by introducing an additional time
coordinate. Completeness of the wave operators for $\wt h, \wt h_o$
was established by Yajima \cite{Y77}.   In \cite{H79}, \cite{K84} it
was shown that $\wt h$ has no singular continuous spectrum.
Moreover, Korotyaev \cite{K84} proved in  that the total number of
embedded eigenvalues on the interval $[0,\o]$, counting
multiplicity, is finite. The case of Schrodinger operators with
time-periodic electric and homogeneous  magnetic field was discussed
in \cite{K80}, \cite{K89}, \cite{Y82}, see also recent papers
\cite{AK19}, \cite{AKS10}, \cite{AK16}, \cite{Ka19}, \cite{M00}.
Moreover, scattering for three body systems was considered in
\cite{K85}, see also \cite{MS04}.

Now we discuss stationary case of discrete multidimensional
Schr\"odinger operators on the cubic lattice $\Z^d, d\ge 2$, when
potentials are real and  do not depend on time.  For Schr\"odinger
operators with decaying potentials on the lattice $\Z^d$, Boutet de
Monvel and Sahbani \cite{BS99} used Mourre's method to prove
completeness of the wave operators, absence of singular continuous
spectrum and local finiteness of eigenvalues away from threshold
energies. Isozaki and Korotyaev \cite{IK12} studied the direct and
the inverse scattering problem as well as trace formulas. Korotyaev
and Moller \cite{KM19} discussed the spectral theory for potentials
$V\in \ell^p, p>1$.   Isozaki and  Morioka \cite{IM14} and
Vesalainen \cite{V14} proved  that the point-spectrum of $H$ on the
interval $(0,2d)$ is absent, see also \cite{AIM16}. An upper bound
on the number of discrete eigenvalues in terms of some norm of
potentials was given by Korotyaev and Sloushch \cite{KS20},
Rozenblum and Solomyak\cite{RoS09}. For closely related problems, we
mention that Parra and Richard \cite{PR18} reproved the results from
\cite{BS99} for periodic graphs.  Finally, scattering on periodic
metric graphs associated with $\Z^d$ was considered by Korotyaev and
Saburova \cite{KS20}.

\section {Regularized determinants }
\setcounter{equation}{0}

\subsection{Preliminary analysis} We present the concepts and facts needed below.
Let $\cH$ be a complex separable Hilbert space.  The class of
bounded and compact operators in $\cH$ we denote by $\cB(\cH)$ and
$\cB_\iy(\cH)$ respectively.  Let $\cB_1(\cH)$ and $\cB_2(\cH)$ be
the trace and the Hilbert-Schmidt class equipped with the norm
$\|\cdot \|_{\cB_1}$ and $ \|\cdot \|_{\cB_2}$, respectively.  If it
is evident which  $\cH$ is meant, we shall write simply $\cB,
\cB_\iy$,... We recall some well-known facts about determinants from
\cite{GK69}.

\no $\bu$ Let $A, B\in \cB$ and $AB, BA\in \cB_1$. Then
\begin{equation}
\label{AB} {\rm Tr}\, AB={\rm Tr}\, BA,
\end{equation}
\begin{equation}
\label{1+AB} \det (I+ AB)=\det (I+BA).
\end{equation}

\no $\bu$ If $A,B\in \cB_1$, then
\begin{equation}
\label{DA1}
\begin{aligned}
& |\det (I+ A)|\le e^{\|A\|_{\cB_1}},
 \\
& |\det (I+ A)-\det (I+ B)|\le
\|A-B\|_{\cB_1}e^{\|A\|_{\cB_1}+\|B\|_{\cB_1}}.
\end{aligned}
\end{equation}
We define the  modified determinant $\det_2(I+A)$ by
\[
\begin{aligned}
\label{DA2}
 \det_2 (I+A) =\det \rt((I+ A)e^{-A}\rt),\\
 \end{aligned}
 \]
and  $I+ A$ is invertible if and only if $\det_2(A)\ne 0$, see
Chapter IV in \cite{GK69}).

\no $\bu$ If $A\in \cB_2$, then (see \cite{GK69})
\[
\lb{DA2x}
\begin{aligned}
|\det_2(I+A)|\le e^{{1\/2}\|A\|_{\cB_2}^2}.
\end{aligned}
\]

\no $\bu$  Let $A, B\in \cB$ and $AB, BA\in \cB_2$. Then
\[
\label{DA2v}
\begin{aligned}
\det_2(I+AB) =\det_2(I+BA).
  \end{aligned}
\]

\no $\bu$  Suppose a function $A(\cdot): \O  \to \cB_1$ is analytic
for a domain $\O \subset {\C}$, and the operator $(I+A(z))^{-1} $ is
bounded  for any $z\in \O$. Then the function $f(z)=\det (I+A(z))$
satisfies
\[
\lb{dF}
 f'(z)=f(z)\Tr \big(I+A(z)\big)^{-1}A'(z)\qqq \forall \ z\in \O.
\]
In order to investigate the determinant $\cD(\l)$  we need a
following lemma.

\begin{lemma}
\lb{TAB} Let operators $A, B\in \cB_2$ act on some Hilbert space
$\cH$. Then
\[
\lb{AB1} e^{A}e^{B}e^{-A-B}-I\in \cB_1,
\]
\[
\lb{AB2} \det \rt( e^{A}e^{-A-B}e^{B}\rt)=\det \rt(
e^{B}e^{A}e^{-A-B}\rt)=1,
\]
\[
\lb{AB3} \det \rt((I+A+B)e^{-A-B}\rt)=\det \rt(
e^{-A}(I+A+B)e^{-B}\rt).
\]
If in addition $I+A$ is invertible and $\cR=(I+A)^{-1}$, then
\[
\lb{AB4} \det \rt((I+A+B)e^{-A-B}\rt)=\det \rt((I+A) e^{-A}\rt) \det
\rt((I+\cR B)e^{-B}\rt),
\]
\end{lemma}

\no{\bf Proof.} Using the Tailor series $e^{z}=1+z+{z^2\/2}+...$ at
$z=A, B, -A-B$ we obtain
$$
e^{A}e^{B}e^{-A-B}=(I+A +A_1)(I+B +B_1)(I-A-B +C_1)=I+G,
$$
where $A_1, B_1, C_1, G$ are some trace operators, which yields
\er{AB1}. We show \er{AB2}. Define the trace class valued function
$F(t)-I$ and the determinant $D(t)$  by
$$
F(t)=e^{tA}e^{tB}e^{-t(A+B)},\qqq D(t)=\det F(t)
$$
for $t\in \R$. From \er{dF} we obtain the derivative
$$
D'(t)=D(t)\Tr(F'(t)F(t)^{-1}).
$$
Using this formula, we get:
$$
D'(t)=D(t)\Tr(F'(t)F(t)^{-1})
$$
$$
=D(t)\Tr \rt(e^{tA}e^{tB}\big(e^{-tB}Ae^{tB}-A\big)e^{-t(A+B)} \rt)
e^{t(A+B)}e^{-tB}e^{-tA}
$$
$$
=D(t)\Tr \rt(\big(e^{-tB}Ae^{tB}-A\big) \rt)=0,
$$
which yields \er{AB2}.  We show \er{AB3}. Using \er{AB2} we obtain
$$
\begin{aligned}
 \det \rt((I+A+B)e^{-A-B}\rt)=\det \rt((I+A+B)e^{-B}e^{-A} \cdot  e^{A}e^{B}e^{-A-B}\rt)
 \\
\det \rt((I+A+B)e^{-B}e^{-A}\rt)  \det \rt( e^{A}e^{B}e^{-A-B}\rt)
=\det \rt(e^{-A}(I+A+B)e^{-B}\rt) \\
 =\det \rt((I+A) e^{-A}\rt) \det
\rt((I+\cR B)e^{-B}\rt),
\end{aligned}
$$
which yields \er{AB3}. If in addition $I+A$ is invertible, then
\er{AB3} gives
$$
\begin{aligned}
 \det \rt((I+A+B)e^{-A-B}\rt)
 =\det \rt((I+A) e^{-A}\rt) \det
\rt((I+\cR B)e^{-B}\rt),
\end{aligned}
$$
which yields \er{AB4}. \BBox

Let $\wt\gB=L^2(\T_\t,\gB)$ be the space of function $v:\T_\t\to
v(t)\in \gB$ for some Banach space $\gB$, which are measurable and
satisfy $\int_0^\t \|v(t)\|_{\gB}^2dt<\iy$. Recall that  operators
$J_1,J_2$ act on $L^2(\T_\t)$ and are given by
\[
\begin{aligned}
\lb{op1}
 (J_1f)(t)=i\int_0^t f(s)ds,\qqq (J_2 f)(t)= i\int_0^\t f(s)ds,
\end{aligned}
\]
 Note that the operator $I+J_1$ is invertible.

\begin{lemma}
\lb{Tre}  i) The operator $\pa=-i\frac{\pa}{\pa t}$ acting  on
$L^2(\T_\t)$ has the resolvent given by
\[
\lb{res1} ( (\pa-\l)^{-1}f)(t)=\int_0^t i
e^{i\l(t-s)}f(s)ds+{iz\/1-z} \int_0^\t e^{i\l(t-s)}f(s)ds, \qq
z=e^{i\t \l},
\]
where $f\in L^2(\T_\t)$ and $\l\in \C\sm \s(\pa)$.

\no ii) Let an operator function $V\in L^2(\T_\t,\cB_2(\cH))$. Then
operators $J_1V$ and $J_2V$ on $\wt\cH$ belong to $\wt\cB_2$ and
satisfy
\[
\lb{res2j}
\begin{aligned}
 \|J_2 V\|_{\wt\cB_2}^2\le \t \int_0^\t \|V(t)\|_{\cB_2}^2 dt\le
\t^2\sup_{t\in [0,\t]} \|V(t)\|_{\cB_2}^2,
\\
\|J_1 V\|_{\wt\cB_2}^2\le \int_0^\t \|V(t)\|_{\cB_2}^2 (\t-t)dt\le
{\t^2\/2}\sup_{t\in [0,\t]} \|V(t)\|_{\cB_2}^2.
\end{aligned}
\]
\end{lemma}

{\bf Proof.} i) We have $(\pa-\l)u=f$, where $u',f\in L^2(\T_\t)$.
Then using $u(\t)=u(0)$  we obtain
$$
\begin{aligned}
u'-i\l u=if,\qq (e^{-i\l t}u)'=i e^{-i\l t}f,\qq e^{-i\l
t}u(t)=u(0)+\int_0^t i e^{-i\l s}f(s)ds
\\
e^{-i\l \t}u(\t)=u(0)+\int_0^\t i e^{-i\l s}f(s)ds,\qq  (e^{-i\l
\t}-1)u(0)=\int_0^\t i e^{-i\l s}f(s)ds,
\end{aligned}
$$
which yields  \er{res1}.

ii) The Gilbert-Schmidt norms of $J_j V, j=1,2$ are
$$
\begin{aligned}
\|J_2 V\|_{\wt\cB_2}^2=\int_0^\t ds\int_0^\t  \|V(t)\|_{\cB_2}^2 dt
=\t \int_0^\t  \|V(t)\|_{\cB_2}^2 dt\le \t^2\sup_{t\in [0,\t]}
\|V(t)\|_{\cB_2}^2,
\\
\|J_1 V\|_{\wt\cB_2}^2=\int_0^\t dt\int_0^t  \|V(s)\|_{\cB_2}^2ds =
\int_0^\t \|V(s)\|_{\cB_2}^2 (\t-s)ds\le {\t^2\/2}\sup_{t\in [0,\t]}
\|V(t)\|_{\cB_2}^2.
\end{aligned}
$$
\BBox

\begin{proposition}
\lb{Tre2}
 Let $h_o$ be a  bounded self-adjoint operator  on the separable
  Hilbert space $\cH$.

\no i) The resolvent $R_o(\l)=(\pa+h_o-\l)^{-1}, \l\in \C\sm \R $ on
$f\in L^2(\T_\t)\ts \cH$ has the form given by
\[
\lb{ddx1}
\begin{aligned}
R_o(\l)f(t)=i e^{it \vp}\int_0^\t \rt(\1_{t-s}+ {e^{i\t
\vp}\/1-e^{i\t \vp}}\rt) e^{-is\vp} f(s)ds,
\\
\vp=\l-h_o,\qq  e^{it \vp}=z a, \qqq z=e^{i\t \l}, \qq
 a=e^{-i\t h_o},
\end{aligned}
\]
where $\1_t=1, t>0$ and $\1_t=0, t<0$.

\no ii) Let, in addition, an operator function $V\in
L^2(\T_\t,\cB_2(\cH))$ and let $c=\int_0^\t \|V(s)\|_{\cB_2}^2ds$.
Then operator $R_o(\l)V$ on $\wt\cH$ belong to $\wt\cB_2$ and
satisfies
\[
\lb{res2} \| R_o(\l)V\|_{\wt\cB_2}^2 \le {2c\/\n(1-e^{-\n \t})},\qqq
\n:=\Im \l>0.
\]
 Moreover, if $\l\in \C_+$ is an eigenvalue of  the operator $\wt
h_o+V$, then
\[
\lb{XX1}  \n(1-e^{-\n \t})\le 2c.
\]
\end{proposition}

{\bf Proof.} The statement i) follows from Lemma \ref{Tre}.

ii) Using \er{ddx1} we present $R_o$ in the form $R_o=R_1+X R_2$,
where
$$
 R_1 (\l)f(t)=\int_0^t i e^{i\l(t-s)\vp}f(s)ds,\qq R_2
(\l)f(t)=\int_0^\t e^{i\l(t-s)\vp}f(s)ds, \qq X={e^{i\t
\vp}\/1-e^{i\t \vp}}.
$$
Consider the case $\n=\Im \l>0 $, the proof for $\n<0$ is similar.
The Gilbert-Schmidt norm of $R_1 V$ is
\[
\label{X1}
\begin{aligned}
\|R_1 V\|_{\wt\cB_2}^2=\int_0^\t e^{-2\n (t-s)}dt\int_0^t
\|V(s)\|_{\cB_2}^2ds =\int_0^\t e^{2\n s}\|V(s)\|_{\cB_2}^2ds
\int_s^\t  e^{-2\n t}dt
\\
={1\/2\n}\int_0^\t \|V(s)\|_{\cB_2}^2 (1-e^{-2\n (\t-s)})ds\le
{c\/2\n}  ,
\end{aligned}
\]
and the Gilbert-Schmidt norm of  $R_2 V$ is
$$
\label{}
\begin{aligned}
\|R_2 V\|_{\wt\cB_2}^2=\int_0^\t\!\! e^{-2\n t}dt\int_0^\t \!\!
 e^{2\n s}\|V(s)\|_{\cB_2}^2ds = {1-e^{-2\n \t}\/2\n} \int_0^\t \!\!
e^{2\n s}\|V(s)\|_{\cB_2}^2ds\le {e^{2\n \t}-1\/2\n}c.
\end{aligned}
$$
Then the estimate $\|X\|\le {e^{-\n \t}\/1-e^{-\n \t}}={1\/e^{\n
\t}-1}$ gives
\[
\label{X2} \|X R_2(\l)V\|_{\wt\cB_2}^2\le  {(e^{2\n \t}-1)\/(e^{\n
\t}-1)^2} {c\/2\n}={(e^{\n \t}+1)\/(e^{\n \t}-1)} {c\/2\n},
\]
since ${(e^{\n \t}+1)\/(e^{\n \t}-1)}=1+{2\/e^{\n \t}-1}\le 1+{2\/\n
\t}$. Then we obtain
\[
\label{X3}
\begin{aligned}
\|R_oV\|_{\wt\cB_2}^2\le
2\|R_1V\|_{\wt\cB_2}^2+2\|R_2V\|_{\wt\cB_2}^2 \le {c\/\n}{2e^{\n
\t}\/(e^{\n \t}-1)}={2c\/\n(1-e^{-\n \t})}.
\end{aligned}
\]
If $\|R_o(\l) V\|_{\wt\cB_2}<1$, then the operator $I+R_o(\l) V$ has
an inverse. Thus from \er{X3} we have that if ${2c\/\n(1-e^{-\n
\t})}<1$, then $\l$ is not an eigenvalue of  the operator $H_o+V$.
Then if $\l$ is an eigenvalue of  the operator $H_o+V$, then $2c\ge
\n(1-e^{-\t\n })$ . \BBox

\subsection  {Determinants}

The operator $\pa=-i{\pa\/\pa t}$ on $L^2(\T_\t)$ has the spectrum
$\s(\pa)=\{\o \Z\}$, where $\o={2\pi\/\t}$.

\begin{lemma}
\label{TPD}

Let $h_o$ be a  bounded self-adjoint operator  on the separable
  Hilbert space $\cH$. Let an operator-valued function $V\in
L^2(\T_\t,\cB_2(\cH))$ and $R_o(\l)=(\pa+h_o-\l)^{-1}, \l\in
\C_\pm$. Then

\no i) Operators $R_o(\l)V$ and $ VR_o(\l)\in \cB_2(\wt\cH)$ for any
$\l\in \C_\pm$; the modified determinant $\cD(\l)=\det
\Big[(I+VR_o)e^{-VR_o}\Big](\l)$ is well defined, analytic in
$\C_\pm$ and satisfies
\begin{equation}
\label{PD1}
 \cD(\lambda)=1+O(1/\n) \quad as \quad  \n:=\Im\l\to
\pm {\infty},\qqq \l\in \C_\pm.
\end{equation}

\no ii) The modified determinant $\cD(\l)$ satisfies
\[
\label{pd22} \cD(\l+\o)=\cD(\l)\qqq \forall \ \l\in \C_\pm,
\]
\begin{equation}
\label{PD1r} {\cD'(\l)\/\cD(\l)}=-\Tr \rt((R_o(\l)V)^2R(\l)\rt),
\end{equation}
\begin{equation}
\label{PD3} \log \cD(\lambda) = -
\sum_{n=2}^{\infty}\frac{(-1)^n}{n}{\rm
Tr}\,\left(R_o(\lambda)V\right)^n,
\end{equation}
where the traces $T_n:=\Tr (VR_o(\l))^n, n\ge 2$ satisfy
\[
\begin{aligned}
\lb{asD1}
 |T_n(\l)|\le  \rt(  {4\/\n} \int_0^\t\|V(s)\|_{\cB_2}^2ds \rt)^{n\/2} \qqq
 \forall \  \n\ge 1/\t.
\end{aligned}
\]

\end{lemma}
{\bf Remark.} Due to \er{PD1} we take the branch of $\log \cD$ so
that $\log \cD(\l )=o(1)$ as $|\Im\l |\to {\infty}$.

{\bf Proof}. i) Lemma \ref{Tre2} gives that $VR_o(\l)\in
\cB_2(\wt\cH)$ for any $\l\in \C_\pm$. We show that the determinant
$\cD$ is well defined.  The Taylor series for the entire function
$e^{-E}$ and the estimate \er{res2} give at $A(\l)=VR_0(\l)$
$$
[(I+A)e^{-A}]=(I+A)(1-A+A^2O(1))=1-A^2+A^2O(1)=I+A^2O(1).
$$
Moreover,  this asymptotics and \er{res2}, \er{DA1} imply \er{PD1}.

 ii) The identities \er{1.c1} and \er{DA2v} yield \er{pd22}.
 Take $|\Im \l|\ge r$ for $r> 0$ large enough. Then from \er{res2}, we
have by the resolvent equation
\begin{equation}
\label{RR0}
 R(\l) =R_0(\l)+
\sum_{n=1}^{\infty}(-1)^n \rt(R_0(\l)V\rt)^{n}R_0(\l),
\end{equation}
where the right-hand side is uniformly convergent on $\{\l\in
\C:|\Im \l|\ge r\}$.  Using  (\ref{dF}) and  \er{AB}, we have  the
following for $\l\in \L$:
\begin{equation}
\lb{D16}
\begin{aligned}
&\cD'(\l)=-\cD(\l)\Tr\rt(e^{A(\l)}(I+A(\l))^{-1}A(\l)A'(\l)e^{-A(\l)}
\rt)
\\
&=-\cD(\l)\Tr (I+A(\l))^{-1}A(\l)A'(\l)
=-\cD(\l)\Tr VR(\l)VR_0^2(\l)\\
&=-\cD(\l)\Tr \rt(R_0(\l)VR(\l)VR_0(\l)\rt) =-\cD(\l)\Tr
\rt((R_0(\l)V)^2R(\l)\rt),
\end{aligned}
\end{equation}
since $R(\l)VR_0(\l)=R_0(\l)VR(\l)$,   which yields \er{PD1r}. Thus
(\ref{RR0}) gives
\begin{equation}
\begin{aligned}
\lb{deD} &(\log \cD(\l))'=-\Tr\, \sum_{n=0}^{\infty}(-1)^n
\rt(R_0(\l)V\rt)^{n+2}R_0(\l).
\end{aligned}
\end{equation}
Then integrating we obtain \er{PD3} since we have the identity
$$
 {d\/d\lambda}\rt(\Tr \rt(VR_0(\lambda)\rt)^{n}\rt)=n
 \Tr\,  \rt(VR_0(\lambda)\rt)^{n}R_0(\lambda).
$$

iii) From $\n \t \ge 1$ we have $1+{1\/\n \t}\le 2$  and using
\er{res2} we obtain  \er{asD1}:
 $$|\Tr T_n|=|\Tr
A^n(\l)|\le \|VR_o(\l)\|_{\wt\cB_2}^n\le \rt( {4\/\n}
\int_0^\t\|V(s)\|_{\cB_2}^2ds  \rt)^{n\/2}.
$$
 \BBox

We are ready to prove the main theorem of this section. Here we
transform the presentation of the  modified determinant $\cD$ in the
forms \er{dd1}, \er{dd4} convenient for us. Via this presentation we
determine the asymptotics \er{PD6} of $\cD(\l)$ in terms of
$z=e^{i\t \l}$ as $\Im \l\to +\iy$. In fact, we determine the Taylor
expansion \er{PD6} of $\p(z)$ in some disk $\{|z|<r\}$

 \begin{theorem}
\lb{TD1} Let $h_o$ be a  bounded self-adjoint operator  on the
separable Hilbert space $\cH$. Let an operator function $V\in
L^2(\T_\t,\cB_2(\cH))$. Then

i) The modified determinant $\cD(\l)=\det
\Big[(I+R_oV)e^{-R_oV}\Big](\l)$ is analytic in $\L$ and the
function $\p(z)=\cD(\l(z))$ is analytic in $\dD$ and has the
following form
\[
\label{dd1} \p(z)=\det \Big[(I+F)e^{-F}\Big](z),\qq \qq
z=e^{i\t\l}\in \dD,
\]
where the operator $F(z)$ acts on $\wt\cH$ and is given by
\[
\begin{aligned}
\lb{dd2} F(z)=F_1+\g(z)F_2,\qqq \g(z)={za\/1-za},\qq a=e^{-i\t h_o},
\\
F_1=J_1 \wt V,\qq  F_2=J_2 \wt V, \qqq \wt
V(t)=e^{ith_o}V(t)e^{-ith_o},
\end{aligned}
\]
where $F_1, F_2\in \cB_2(\wt\cH)$. Moreover,
 the operator $I+F_1$ is  invertible and if $\cR_1 =(I+F_1)^{-1}$, then
\[
\begin{aligned}
\lb{dd4} \p(z)=\det \Big[(I+\cR_1 \g F_2)e^{-\g F_2}\Big](z),\qqq
|z|<1.
\end{aligned}
\]
\no iii) The function $\log \p(z)$ is analytic in $\dD_{r}$ for some
$r>0$ and has the following form
\[
\begin{aligned}
\lb{dd5} \log \p(z)=\sum_{n=1}^{\infty} \frac{(-1)^n}{n}\Tr\,
F_1\left(\cR_1 \g(z)F_2\right)^n.
\end{aligned}
\]
Moreover, it has the following Taylor series
\[
\label{PD6} \log \p(z)=\p_1z+\p_2z^2+\p_3z^3+......, \qqq \as \qq
|z|<r,
\]
where
\[
\label{PD7} \p_1=-\Tr F_1 \cR_1 a F_2, \qqq \p_2= -\Tr F_1 \cR_1
a^2F_2+{1\/2}\Tr F_1 (\cR_1  a F_2)^2,....
\]

\end{theorem}

\noindent {\bf Proof.} Due to Lemma \ref{TPD} an operator
$R_o(\l)V\in \cB_2(\wt\cH)$ for any $\l\in G$. Recall that $z=e^{i\t
\l}$, where $\l\in \L$ and $\1_t=1, t\ge 0$ and $\1_t=0, t<0$. Due
to \er{ddx1} the operator $(\pa+h_o-\l)^{-1}V, \l\in \C_\pm$ on
$\wt\cH=L^2(\T_\t,\cH)$ has the form given by
\[
\lb{ddx1v}
\begin{aligned}
(R_o(\l)Vf)(t)   =i e^{it\vp}\int_0^\t \rt(\1_{t-s}+ \g(z)\rt)
e^{-is\vp} V(s)f(s)ds =bF(z) b^{-1} f,
\\
\vp=\l-h_o,\qq  e^{i\t \vp}=z a, \qqq z=e^{i\t \l}, \qq
 a=e^{-i\t h_o},
\end{aligned}
\]
where $f\in \wt\cH$ and $b=e^{it\vp}$ is a multiplication operator
$e^{it\vp}f(t)$ in $\wt\cH$. Then from \er{DA2v} we obtain for
$\p(z)=\cD(\l(z))$:
$$
  \p(z)=\det \Big[(I+R_oV)e^{-R_oV}\Big](\l(z))=\det
\Big[(I+F)e^{-F}\Big](z).
$$
We have the decomposition $F=F_1+\g F_2$, where due to \er{res2j}
the operators $F_1, F_2\in \cB_2(\wt\cH)$ and  the operator $F_1$ is
invertible, as the Volterra operator. Thus due to Lemma \ref{TAB} we
obtain
\[
\begin{aligned}
\lb{dd4x} \p(z)=\det \Big[(I+F_1)e^{-F_1}\Big]\cD_2(z),\qqq
\cD_2(z):=\det \Big[(I+\cR \g(z) F_2)e^{-\g(z)F_2}\Big].
\end{aligned}
\]
Recall that the operator $\wt V(t)=e^{ith_o}V(t)e^{-ith_o}$, where
$V(t)\in \cB_2(\cH)$.  Due to \er{res2j} the operator $F_2\in
\cB_2(\wt\cH)$ and if $z\to 0$ then we obtain
$$
\|\g(z)F_2\|_{\cB_2}\le{|z|\/1-|z|}\|F_2\|_{\cB_2}=O(|z|)\|F_2\|_{\cB_2},
$$
which yields $\cD_2(z)\to 1$. From \er{PD1} we have $\p(z)\to 1$.
From \er{dd4x} we obtain
$$
\p(z)=\det \Big[(I+F_1)e^{-F_1}\Big] \cD_2(z)\to 1, \qq \cD_2(z)\to
1,
$$
which yields $\det \Big[(I+F_1)e^{-F_1}\Big]=1$ and we get \er{dd4}.

We show \er{dd5}. Estimates in \er{res2j} yield
\[
\lb{oo1} \|F_1\|_{\cB_2}\le C_1,\qqq \|F_2\|_{\cB_2}\le C_2,\qq
\|\cR\|\le C_o,
\]
for some constants $C_1, C_2, C_o$. Let $A=\g F_2$ and let
$D(t)=\det \Big[(I+\cR tA)e^{-t A}\Big], t\in \R$. From \er{dF},
\er{oo1} and $\cR-I=-F_1\cR$ we obtain
\[
\begin{aligned}
\lb{dd4c}
& D'(t)/D(t)=\Tr  \rt[\rt( \cR A - A\rt) e^{-t A}\rt]e^{t
A}(I+\cR tA)^{-1}
\\
&=\Tr ( \cR-I)A(I+\cR tA)^{-1}=-\Tr F_1\cR A(I+\cR tA)^{-1}=\Tr
F_1\sum_{n\ge 0} (-\cR A)^{n+1} t^n
\end{aligned}
\]
and the integration yields \er{dd5} for $z$ small enough.

Let $\z=z a,\ a=e^{-i\t h_o}$.  We have
$\g(z)={\z\/1-\z}=\z+\z^2+\z^3+\z^4+...,$ and then
\[
\g(z)F_2 =(\z+\z^2+\z^3+\z^4+...)F_2, \qq
\]

From \er{dF}, \er{oo1} and \er{dd5}  we obtain for the term with
$n=1$ and $n=2$:
\[
\begin{aligned}
\lb{aa1} & -\Tr\, F_1\cR \g(z)F_2=-\Tr\, F_1\cR (za+z^2a^2+...)F_2,
\\
& \Tr\, F_1(\cR F_2(z))^2=\Tr\, F_1\rt(\cR
(za+z^2a^2+...)F_2\rt)^2=z^2\Tr\, F_1(\cR  a F_2)^2+O(z^3).\textbf{}
\end{aligned}
\]
Collecting asymptotics from \er{aa1} we obtain \er{PD7}. \BBox

\section{Proof of main theorems}
\setcounter{equation}{0}

\subsection  {Laplacian on the lattice }
 We need results about the resolvent
$r_o(\l)=(\D-\l)^{-1}$ on $\ell^{2}(\Z^d)$ from \cite{KM19}:

\begin{theorem}
\lb{T2HS} Let $d\geq 3$. Let $u,v\in \ell^{2p}(\Z^d)$ with
 $1\leq p < \ca {6\/5} & \textup{if} \ \ d=3\\
  \frac{3d}{2d+1} & \textup{if}  \  \ d\ge 4 \ac$.
 Then the operator-valued function $\gf\colon \C\sm [0,2d]\to \cB_2$, defined
by
\[
\gf(\l) := u (\D -\l)^{-1}v
\]
is analytic and H\"older continuous up to the boundary and satisfies
for all $\l\in \C\sm [0,2d]$:
\begin{equation}
\lb{efuv} \|\gf(\l)\|_{\cB_2}\le C_* \|u\|_{2p}\|v\|_{2p},
\end{equation}
where
\[
\label{Cpd}
\begin{aligned}
 C_*=p^{d(p-1)\/2 p}+c_d (3+2 c )^{d-{d\/p}},\qq
c_d=\ca 16\\
           4\\
           {14\cdot 2^{d\/4}\/d-4}\ac,\
 c=\ca {6(p-1)\/6-5p}\ & if \ d=3\\
      \rt({5p-1\/4-3p}\rt)^{5p-4\/4(p-1)}\ & if \ d=4\\
      {3d(p-1)\/3d-(2d+1)p}\ & if \ d\ge 5\ac .
\end{aligned}
\]
Moreover,  we have
\[
\begin{aligned}
\label{y21} \|\gf(\l)-\gf(\m)\|_{\cB_2}\le C_\alpha |\l-\m|^\a
\|u\|_{2p}\|v\|_{2p},\qqq \forall \ \ \qq \l,\m\in \ol \C_\pm,
\end{aligned}
\]
where  $\a, C_\a$ are some positive constants.
\end{theorem}

Below we need a simple corollary

\begin{corollary}
\lb{Tsp1}

Under the conditions of Theorem \ref{T2HS} the operator-valued
function $\gf: \C\sm [0,2d]\to \cB_2$  satisfies
\[
\label{y11}
\begin{aligned}
 \|\gf(\l)\|_{\cB_2}\le C_*{\|u\|_{2p}\|v\|_{2p}\/\max \{1,\gr(\l)\}},
 \qqq \forall \ \ \l\in \C\sm [0,2d],
\end{aligned}
\]
where $\gr(\l)=\dist\{\l,\s(\D)\}$, and
 and the constant $C_*$ is  defined by \er{Cpd}.

\end{corollary}

\no {\bf Proof.} Let $\l\in \C\sm [0,2d]$. If $\gr(\l)\le 1$, then
from \er{efuv} we obtain \er{y11}. If $\gr(\l)\ge 1$, then we have
$$
\|\gf(\l)\|_{\cB_2}\le \|u\|_{2}\|v\|_{2}\|r_o(\l)\| \le
{\|u\|_{2p}\|v\|_{2p}/\gr(\l)},
$$
which yields \er{y11}.
 \BBox

In order to prove main theorem we need a simple estimate.

\begin{lemma}
\lb{Tg1} Consider a function   $g(a)={1\/a} -{e^{-i\vk a}\/\sin a}$
in a domain $\gS_+:=\{a\in \ol\C_+:|\Re a|\le  {3\pi\/4}\}$ for some
parameter $\vk\in [-3,1]$. Then
\[
\lb{g2} \max_{\gS_+} |g|\le C_g:={4\/3\pi}+{5+3\pi\/4}\sqrt2.
\]

\end{lemma}
{\bf Proof.}
 We have a simple decomposition
\[
\lb{cf1t}
\begin{aligned}
g={1\/a} -{e^{-i\vk a}\/\sin a}=\gs+f, \ \ \where\ \
\gs={1\/a}-{1\/\sin a},\qq f={1-e^{-i\vk a}\/\sin a}
\end{aligned}
\]
and the functions $\gs, f$ is analytic in the strip $\gS=\{|\Re a|<
{3\pi\/4}\}$. By the maximum principle, the function $\gs$ in the
strip $\gS$ has maximum on the lines $\Re a=b:={3\pi\/4}$, which
yields
\[
\lb{gs1x} \max_\gS |\gs|=\max_{t\in \R} |\gs(b+it)|\le \max_{t\in
\R} {1\/|\sin (b+it)|}+ \max_{t\in \R}{1\/|b+it|} = {\sqrt
2}+{4\/3\pi}.
\]
Consider the function $f$ in the half strip $\gS_+$. We have
\[
\lb{gs2}
\begin{aligned}
\max_{\gS_+} |f|=\max \{f_\pm, f_o\}, \qq   f_\pm=\max_{t\ge 0}
|f(\pm b+it)|,\qq   f_o=\max_{a\in [-b,b]} |f(a)|.
\end{aligned}
\]
Let $a=\pm b+it, \ t\ge 0$. We obtain
\[
\lb{gs3}
\begin{aligned}
|f(a)|={|1-e^{-i\vk a}|\/|\sin a|} \le {1\/|\sin a|}+{|e^{-i\vk
a}|\/|\sin a|}\le \sqrt 2+{e^{\vk t}\/|\sin a|},
\\
{e^{\vk t}\/|\sin a|}={2e^{\vk t}\/|e^{ia}-e^{-ia}|}={2e^{(\vk-1)
t}\/|e^{i2a}-1|}={2e^{(\vk-1) t}\/|ie^{-2t}+1|}\le 2,
\end{aligned}
\]
since $e^{i2a}=e^{\pm i{3\pi\/2}}e^{-2t}=\mp ie^{-2t}$. This yields
$f_\pm\le \sqrt 2+2$.

Let $-b\le a\le b$. Then we have $|f(a)|={|1-e^{-i\vk a}|\/|\sin
a|}= {2|\sin{\vk a\/2}|\/|\sin a|}$ and we obtain
$$
\begin{aligned}
|f(a)|= {2|\sin{\vk a\/2}|\/|\sin a|} \le {|\vk a|\/|a{\sin
{\pi\/4}\/{\pi\/4}}|}=|\vk| {\pi\/4}\sqrt2\le {3\pi\/4}\sqrt2,\qqq
if \qq |a|\le {\pi\/4},
\\
|f(a)|= {2|\sin{\vk a\/2}|\/|\sin a|} \le {2\/\sin
{\pi\/4}}=2\sqrt2,\qqq if \qq {\pi\/4}\le |a|\le b.
\end{aligned}
$$
This yields $f_o\le {3\pi\/4}\sqrt2$ and then $|g|\le {\sqrt
2}+{4\/3\pi}+{3\pi\/4}\sqrt2$.
 \BBox

\subsection  {Proof of main theorems}

Consider the operator $\wt h=\wt h_o +V$ on
$\wt\cH=L^2(\T_\t,\ell^2(\Z^d))$, where  $\wt h_o=\pa +\D$ is the
free operator. Due to the factorization $ V=vq$, we define the
operator-valued function $\gF(\l)$ on $\wt\cH$ by
\[
\lb{fF1}
 \gF(\l)=qR_ov,\qqq  \l \in \C_\pm,\qq \where \qq R_o(\l)=(\wt
 h_o-\l)^{-1},\qq
q=|V|^{1\/2},\qq v=qe^{i\arg V}.
\]

\smallskip
\begin{theorem}
\lb{Ty1} Let  $V$ satisfy Condition V and the operator $\gF(\l), \l
\in \C_\pm$ be defined by \er{fF1}. Then
\[
\label{Y00}
\begin{aligned}
\gF(\l), \qq  VR_o(\l)\in \wt\cB_2:=\cB_2(\wt \cH), \qqq \forall \ \
\l\in \C_\pm.
\end{aligned}
\]
 Moreover, if we define $\|V\|_{p,1}=\int_0^\t \|V(t)\|_{\ell^p(\Z^d)}dt$, then
  the operator-valued function $\gF: \C_\pm\to \wt\cB_2$ is analytic and
H\"older continuous up to the boundary  and satisfies
\[
\label{F01}
\begin{aligned}
 \|\gF(\l)\|_{\wt\cB_2}\le C_\bu \|V\|_{p,1},\qqq \forall \ \ \l\in \C_\pm,
  \\
C_\bu={1\/\sqrt2}+ C_\t \rt(C_g+{\sqrt 2\/ \sqrt {\pi \t}}
+{C_*\/\t}\rt),\qq C_\t=1+{\t d\/\pi},\qq
C_g={4\/3\pi}+{5+3\pi\/4}\sqrt2,
\end{aligned}
\]
and
\[
\begin{aligned}
\label{F02} \|\gF(\l)-\gF(\m)\|_{\wt\cB_2}\le C_{1,\a} |\l-\m|^\a
\|V\|_{p,1},\qqq \forall \ \ \qq \l,\m\in \ol \C_\pm,
\end{aligned}
\]
where  $\a, C_{1,\a}$ are some positive constants and the constant
$C_*$ is  defined  by \er{Cpd}.
\end{theorem}

 {\bf Proof.}  The results in \er{Y00} have been proved in Lemma
 \ref{Tre2}.

Due to \er{i1} it is enough to prove \er{F01} only for the case
$\l\in \L=[0,\o]\ts i\R_+$.

Using \er{ddx1} we present $\gF f$, where $f\in \wt\cH, \l\in \C_+$
in the form given by
\[
\lb{wR1}
\begin{aligned}
& (\gF(\l)f)(t)= i q(t)e^{it\vp}\int_0^\t \rt(\1_{t-s}+
\g\rt)e^{-is\vp}v(s) f(s)ds=(\gF_1(\l)f)(t)+(\gF_2(\l)f)(t),
\\
& (\gF_1(\l)f)(t)=i q(t)\!\! \int_0^t \!\! e^{i(t-s)(\l-\D)}v(s)
f(s)ds,\ \ (\gF_2(\l)f)(t)=q(t)  \!\!\int_0^\t \!\! \vt(A,\vk) v(s)
f(s)ds,
\end{aligned}
\]
where
\[
\lb{wR2}
\begin{aligned}
& \vp=\l-\D, \qq A={\t \vp\/2}= {\t (\l-\D)\/2},\qq  \g={e^{i\t
\vp}\/1-e^{i\t \vp}}=-{e^{iA}\/2i\sin A},
\\
& \vk={2(s-t)\/\t}-1\in [-3,1],\qq \vt(\a,\vk)=-{e^{-i\vk\a}\/2\sin
\a}, \qq \a\in \R, \qq \o={2\pi\/\t}.
\end{aligned}
\]
 The first term,  the  operator-valued function $\gF_1:\C\to \wt\cB_2$ is entire
and satisfies
\[
\lb{wR3}
\begin{aligned}
& \|\gF_1\|_{\wt\cB_2}^2=\int_0^\t \!\! dt\int_0^t  \|q(t)
e^{i(t-s)(\l-\D)}v(s)\|_{\cB_2}^2ds\le \int_0^\t \!\! dt\int_0^t
\|q(t)\|_{\cB_4}^2 \|q(s)\|_{\cB_4}^2 e^{-2(t-s)\Im \l}ds
\\
& \le \int_0^\t \|q(t)\|_4^2 dt\int_0^t \|q(s)\|_4^2 ds=
{1\/2}\rt(\int_0^\t \|V(t)\|_2 dt\rt)^2= {1\/2} \|V\|_{2,1}^2,\qq
\forall \ \Im \l\ge 0,
\end{aligned}
\]
where $\|q(t)\|_a=\|q(t)\|_{\cB_a}$ is the norm in
$\ell^a(\Z^d),a\ge1 $ and $\|q(t)\|_4^2=\|V(t)\|_2$.

Define the interval $I(\o)=[\l_o-{\o\/2}, \l_o+{\o\/2}]$ for any
fixed $\l_o\in [0,2d]\cap [0,\o]$. Define  $\c$ by

\[
\lb{cf}
\begin{aligned}
\c=\sum_{j} \c_j,\qq  \c_0(\m)=\ca 1, \ \   \m\in I(\o)
 \\
             0,\ \  \m\in \R\sm I(\o) \ac,\qq  \c_j=\c_0(\cdot-\o
             j),\qq j\in \Z.
\end{aligned}
\]
Note that the sum $\c(\D)=\sum_{j} \c_j(\D)$ is finite and the
number of the function $\c_j(\D)\ne 0$ is less than $1+{\t d\/\pi}$,
since $\s(\D)=[0,2d]$.

Consider the operator-valued function $\gF_2$. We rewrite one in the
form
\[
\lb{cf1}
\begin{aligned}
& (\gF_{2}(\l)f)(t)=q(t)\int_0^\t \vt(A,k)\c(\D) v(s) f(s)d
=\sum_{j}(F_{j}(\l)f)(t),
\\
&  (F_{j}(\l)f)(t)=  q(t)\int_0^\t \vt(A,k)\c_j(\D) v(s) f(s)ds.
\end{aligned}
\]
 We consider smoothness and estimates  of $F_{0}(\l)$.
The proof for other $F_j, j\ne 0$ is similar. We
 present $\vt(A,k)$ in the following form
\[
\vt(a,\vk)=-{1\/2a}+{g(a)\/2},\qq    g(a):={1\/a}-{e^{-i\vk a}\/\sin
a}.
\]
Thus we obtain
\[
\lb{cf2}
\begin{aligned}
 F_{0}(\l)f(t)=q(t)\int_0^\t \vt(A,\vk)\c_0(\D) v(s) f(s)ds
\\
 =
{q(t)\/2}\int_0^\t \rt(-{1\/A}+{g(A)}\rt)\c_0(\D) v(s) f(s)ds,
\end{aligned}
\]
 which yields $ F_{0}=F_{0r}+F_{0g}$, where
\[
\lb{cf2zz}
\begin{aligned}
 F_{0r}(\l)f(t)=  {q(t)\/\t}\int_0^\t r_o(\l)\c_0(\D) v(s)f(s)ds,
 \\
F_{0g}(\l)f(t)= {q(t)\/2}\int_0^\t g(A)\c_0(\D) v(s) f(s)ds,
\end{aligned}
\]
and $r_o(\l)=(\D-\l)^{-1}$. In order to consider  the
operator-valued function $F_{0g}:\C\to \wt\cB_2$ we need to discuss
the function $g(A)\c_0(\D)$.   We have
$$
 \textstyle A= {\t\/2} (\l-\D)=\z+A_0,\qqq \where \qq A_o= {\t\/2}
(\l_o-\D),\qq \z={\t \/2}(\l-\l_o)
$$
and
$$
\begin{aligned}
 \textstyle \|A_o \c_0(\D)\|\le {\t\o\/4}={\pi\/2},\qqq
 \ca \ |\z|<  {\t\o\/4}={\pi\/2}\qq if \qq |\l-\l_o|<{\o\/2}
 \\
  \ |\z|< {\t\o\/8}={\pi\/4} \qq if \qq  |\l-\l_o|<{\o\/4} \ac .
\end{aligned}
$$
Thus we have
$$
F_{0g}(\l)f(t)= {q(t)\/2}\int_0^\t g(A_o+\z)\c_0(\D) v(s) f(s)ds
$$
where $g(a)$ is analytic in the domain  $\gS_+:=\{a\in \ol\C_+:|\Re
a|\le  {3\pi\/4}\}$ and  due to \er{g2} it satisfies
$\|g(A)\c_0(\D)\|\le C_g={4\/3\pi}+{5+3\pi\/4}\sqrt2$. Then we
obtain
\[
\lb{Fg}
\begin{aligned}
&\|F_{0g}(\l)\|_{\wt\cB_2}^2=\int_0^\t dt\int_0^\t  \|q_t
{g(A)\/2}\c_0(\D)v_s\|_{\cB_2}^2ds
 \le
 {C_g^2\/2}\int_0^\t dt\int_0^\t \|q(t)\|_{\cB_4}^2 \|q(s)\|_{\cB_4}^2 ds
\\
&= {C_g^2\/2}\int_0^\t \|q(t)\|_4^2 dt\int_0^\t \|q(s)\|_4^2 ds=
{C_g^2\/2} \|V\|_{2,1}^2.
\end{aligned}
\]
Consider  the  operator-valued function $F_{0r}:\L\to \wt\cB_2$. We
have $F_{0r}=G_{0}+G_{1}$, where
$$
\begin{aligned}
 (G_{0}(\l)f)(t)= { q(t)\/\t}\int_0^\t r_o(\l) v(s) f(s)ds,\ \
(G_{1}(\l)f)(t)= { q(t)\/\t}\int_0^\t r_o(\l)(\c_0(\D)-1)v(s)
f(s)ds.
\end{aligned}
$$
Due to Theorem \ref{T2HS} the operator-valued function $G_{0}: \C_+
\to \cB_2$ is analytic and H\"older continuous up to the boundary
and via  \er{efuv} satisfies
\[
\lb{Fg2}
\begin{aligned}
\|G_{0}(\l)\|_{\wt\cB_2}^2={1\/\t^2}\int_0^\t dt\int_0^\t  \|q(t)
r_o(\l)v(s)\|_{\cB_2}^2ds
 \le
 {C_*^2\/\t^2}\int_0^\t dt\int_0^\t \|q(t)\|_{2p}^2 \|q(s)\|_{2p}^2 ds
\\
= {C_*^2\/\t^2}\int_0^\t \|V(t)\|_p dt\int_0^\t \|V(s)\|_p ds=
{C_*^2\/\t^2} \|V\|_{p,1}^2.
\end{aligned}
\]
The operator-valued function $r_o(\l)(\c_0(\D)-1): I({\o\/2})\ts
\R_+ \to \cB$ is analytic and H\"older continuous up to the boundary
$I({\o\/2})\ss I({\o})$  and satisfies $\|r_o(\l)(\c_0(\D)-1)\|\le
{4\/\o}$. Moreover, we have
\[
\lb{Fg23}
\begin{aligned}
\|G_{1}(\l)\|_{\wt\cB_2}^2={1\/\t^2}\int_0^\t dt\int_0^\t  \|q(t)
r_o(\l)(\c_0(\D)-1)v(s)\|_{\cB_2}^2ds
\\
 \le
 {2\/\pi \t}\int_0^\t dt\int_0^\t \|q(t)\|_{4}^2 \|q(s)\|_{4}^2 ds
= {2\/\pi \t}\int_0^\t \|V(t)\|_2 dt\int_0^\t \|V(s)\|_2 ds= {2\/\pi
\t}\|V\|_{2,1}^2.
\end{aligned}
\]
Collecting all estimates we obtain
$$
\|F_0(\l)\|_{\wt\cB_2}\le \rt(C_g+{\sqrt 2\/ \sqrt {\pi \t}}
\rt)\|V\|_{2,1}+{C_*\/\t}\|V\|_{p,1}.
$$
Recall that the sum $\c(\D)=\sum_{j} \c_j(\D)$ is finite and the
number of the function $\c_j(\D)\ne 0$ is less than $C_\t=1+{\t
d\/\pi}$, since $\s(\D)=[0,2d]$. Thus we have
$$
\|\sum_{j}(F_{j}(\l)\|_{\wt\cB_2}\le C_\t \rt(C_g+{\sqrt 2\/ \sqrt
{\pi \t}} \rt)\|V\|_{2,1}+C_\t{C_*\/\t}\|V\|_{p,1}
$$
and jointly with \er{wR3} we obtain \er{F01}.
\BBox

{\bf Proof of Theorem \ref{T1}}  Let $V$ satisfy \er{V} and a
constant $C_*$ be defined by \er{Cpd}.

i) Due to Theorem \ref{Ty1}  the operator-valued function $\gF\colon
\C_\pm \to \cB_2(\ell^2(\Z^d))$, defined by \er{dgF}  is analytic
and H\"older continuous up to the boundary. Moreover,  it satisfies
\er{D1}.

ii) From i) we deduce that the modified determinant $\cD(\l)=\det
\Big[(I+\gF)e^{-\gF}\Big](\l)$ is analytic in $\C_\pm$, is H\"older
up to the boundary. Identity \er{i1} at $V=0$ implies \er{D2}, i.e.,
$\cD(\l+\o)=\cD(\l)$ for all $\l\in \C_\pm$. Asymptotics \er{PD1}
yields \er{D3}. From \er{D1} and \er{DA2x} we obtain \er{D4}.
Moreover, estimate \er{ee} has been proved in  Proposition
\ref{Tre2}.
 \BBox

{\bf Proof of Theorem \ref{T2}.} From Theorem \ref{T1} we obtain
that the function $\p\in \mH_\iy(\dD)$ and is H\"older up to the
boundary and satisfies \er{D1}. The zeros $\{z_j\}_{j=1}^N$ of  $\p$
in $\dD$ satisfy $  \sum _{j=1}^N (1-|z_j|)<\iy$.
 Moreover, the function $\log \p(z)$
is analytic in $\dD_{r_0}$ and has the Taylor series
$$
 \log \p(z)=\p_1z+\p_2z^2+\p_3z^3 +......, \qqq \as \qq |z|<r_o.
$$
where coefficients $\p_n$ are given by Theorem \ref{TD1}. \BBox

We recall the standard facts about the canonical factorization of
functions  from Hardy space, see e.g. \cite{G81}, \cite{Ko98} and in
the needed specific  form for us from \cite{KL18}.

\begin{theorem}
\lb{Ta} Let a function $\p\in \mH_\iy(\dD)$ and be H\"older up to
the boundary. Then there exists a singular measure $\gm\ge 0$ on
$[0,2\pi]$, such that $\p$ has a canonical factorization for all
$|z|<1$ given by
\[
\lb{cfDa}
\begin{aligned}
& \p(z)=B(z)e^{\P(z)},\qqq
\P(z)={1\/2\pi}\int_{-\pi}^{\pi}{e^{it}+z\/e^{it}-z}d\m(t),
 \end{aligned}
\]
 where the measure $d\m(t)=\log |\p(e^{it})|dt-d\gm(t)$ and
  $\log |\p(e^{it}) |\in L^1(\T)$ and
the measure $\gm$ satisfies $ \supp \gm\ss \{t\in [0,2\pi]:
\p(e^{it})=0\}$. Moreover, $\P$ has   the Taylor series
\[
\lb{cfDaxx} \P(z)={\m(\T)\/2\pi}+\m_1z+ \m_2z^2+\m_3z^3+\m_4z^4+...,
\]
in some disk $\{|z|<r\}, r>0$, where
$$
\m(\T)=\int_0^{2\pi}d\m(t)=\int_0^{2\pi}\log
|f(e^{it})|dt-\gm(\T),\qqq \m_n={1\/\pi}\int_0^{2\pi}e^{-int}
d\m(t),\qqq n\in \N.
$$

\end{theorem}

We are ready to describe the function $\p$.

{\bf Proof of Corollary \ref{T3}.} The proof follows from Theorem
\ref{T2} and Theorem \ref{Ta}. \BBox

We describe trace formulae.

{\bf Proof of Theorem \ref{T4}.} Using \er{PD1r} and the identity
$R=R_o-R_oVR_o+R_oVR_oVR$  we obtain
$$
\textstyle {\cD'(\l)\/\cD(\l)}=-\Tr \big((R_oV)^2R\big)(\l)=-\Tr
\big(R-R_o+R_oVR_o) \big)(\l).
$$
Then differentiation of $\p(z)=\cD(\l(z))$ in \er{cfD} yields
$$
{\p'(z)\/\p(z)}=\sum {(1-|z_j|^2)\/(z-z_j)(1-\ol
z_jz)}+{1\/\pi}\int_{0}^{2\pi}{e^{it}d\m(t)\/(e^{it}-z)^2}={\cD'(\l(z))\/\cD(\l(z))}\l'(z),
$$
where $\l'(z)={1\/z'(\l)}={1\/i\t z}$. Thus collecting the two last
identities we get \er{trR}.

Due to the canonical representation \er{cfD},  the function
${\p(z)\/B(z)} $
 has no zeros in the disc $\dD$ and $\P(z)=\log
{\p(z)\/B(z)}$ (see \er{cfDa}) satisfies
\[
\lb{Si}
\P(z)={1\/2\pi}\int_{-\pi}^{\pi}{e^{it}+z\/e^{it}-z}d\m(t),\qq
z\in\dD,
\]
where the measure $d\m=\log |f(e^{it})|dt-d\gm(t)$. In order to show
\er{t1}--\er{t3} we need the asymptotics of the Schwatz integral
$\P(z)$  as $z\to 0$ from \er{cfDaxx}:
\[
\lb{asm}
{1\/2\pi}\int_{-\pi}^{\pi}{e^{it}+z\/e^{it}-z}d\m(t)={\m(\T)\/2\pi}+\m_1z+
\m_2z^2+\m_3z^3+\m_4z^4+...\qqq as \qqq |z|<1,
\]
where
$$
\m(\T)=\int_0^{2\pi}d\m(t)=\int_0^{2\pi}\log
|f(e^{it})|dt-\n(\T),\qqq \m_n={1\/\pi}\int_0^{2\pi}e^{-int}
d\m(t),\qqq n\in \Z.
$$

We have the identity $ \log \p(z)=\log B(z)+
{1\/2\pi}\int_{-\pi}^{\pi}{e^{it}+z\/e^{it}-z}d\m(t)$ for all $z\in
\dD_{r_0}$. Combining asymptotics \er{D32}, \er{B6d} and \er{asm} we
obtain \er{t1}-\er{t2}. In particular, we have $-\log B(0)=\sum_1^N
\Im \t\l_j\ge 0$ and ${\m(\T)\/2\pi}\ge 0$  and
 \er{t3}. \BBox

{\bf Proof of Corollary \ref{T5}.}
 Let a potential $V$ satisfy \er{V}. Then substituting estimate \er{D4}
 into \er{t1} we obtain
$$
 {\n(\T)\/2\pi}+\t\sum_{\l\in \L}\Im
\l_j={1\/2\pi}\int_{-\pi}^{\pi}\log |\p(e^{it})|dt\le
{1\/2\pi}\int_{-\pi}^{\pi}{C_\bu^2\/2}\|V\|_{p,2}^2dt={C_\bu^2\/2}\|V\|_{p,2}^2,
$$
which yields \er{e1}. \BBox

\medskip

\no\textbf{Acknowledgments.} \footnotesize EK study was partly
supported by the RSF grant No 18-11-00032.

\end{document}